\documentclass{nature}

\usepackage[utf8]{inputenc}
\usepackage[T1]{fontenc}
\usepackage{microtype}
\usepackage{xcolor}
\usepackage{graphicx}
\usepackage{booktabs}
\usepackage{amsmath, amssymb, mathtools}
\usepackage{bm}
\usepackage[super,comma,sort&compress]{natbib}
\usepackage[labelfont=bf, font=small]{caption}
\usepackage[hidelinks]{hyperref}
\usepackage{enumitem}
\usepackage[section]{placeins}
\usepackage{float}
\usepackage{setspace}
\newcommand{\F}{\mathbf{F}}
\newcommand{\rr}{\mathbf{r}}

\newcommand{\xxi}{\boldsymbol{\xi}}
\newcommand{\tht}{\theta}
\newcommand{\score}{S_\theta}

\title{Path-probability gradients for optimizing force fields
against dynamical observables}

\author{Zhongyao Wang$^{1,2,3,*}$, Lun Song$^{1,4,*}$, Yakun Zeng$^{5,*}$, Xingao Gong$^{6}$, Rui Shi$^{5,\dagger}$, Mao Su$^{1,7,\dagger}$}

\begin{document}
\maketitle

\begin{affiliations}
 \item Shanghai Artificial Intelligence Laboratory, Shanghai 200232, China
 \item College of Intelligent Robotics and Advanced Manufacturing, Fudan University, Shanghai 200433, China
 \item Shanghai Innovation Institute, Shanghai 200231, China
 \item Huazhong University of Science and Technology, Wuhan 430074, China
 \item School of Physics, Zhejiang University, Hangzhou 310058, China
 \item Key Laboratory of Computational Physical Sciences (Ministry of Education), Institute of Computational Physical Sciences, State Key Laboratory of Surface Physics, and Department of Physics, Fudan University, Shanghai 200433, China
 \item Shenzhen Institute of Advanced Technology, Chinese Academy of Sciences, Shenzhen 518055, China
\end{affiliations}

\noindent$^*$These authors contributed equally.
$^\dagger$Correspondence: ruishi@zju.edu.cn, sumao@pjlab.org.cn.

\begin{abstract}
Machine learning force fields inherit the errors of the electronic-structure methods used to generate their training data, and refining them against experimental measurements provides a direct route beyond this limit. 
However, such refinement remains difficult for dynamical observables because differentiating chaotic molecular trajectories yields unstable gradients.
Here, we introduce PathGrad, which differentiates Langevin path probabilities instead of the trajectories themselves. 
The resulting gradient estimator is accumulated on the fly from thermal noise and force derivatives, and its variance remains stable as the trajectory length increases. 
We validated the gradient estimator against finite-difference calculations and used it to fine-tune molten-copper and water potentials against experimental diffusion coefficients together with structural constraints.
For water, jointly optimizing diffusion and the oxygen–oxygen radial distribution function also improves the predictions of shear viscosity and rotational relaxation, neither of which is included in the optimization objective. 
Alignment at a single temperature also improves the apparent activation energies of multiple dynamical observables, bringing their temperature dependences closer to experiments.
Moreover, the optimized potential exhibits smaller force errors relative to an independent, higher-level quantum-chemical reference, indicating that alignment with experiments improves the fidelity of the potential energy surface beyond the fitted targets. 
PathGrad thus establishes a general framework for incorporating experimental observables directly into force field optimizations. 
\end{abstract}

\clearpage

\section{Introduction}

\label{sec:intro}

The predictive accuracy of molecular dynamics (MD) simulation depends on the quality of the force field. Classical empirical force fields are often constructed from physically motivated functional forms and fitted to a combination of quantum-mechanical calculations and experimental observables. Machine learning force fields greatly increase the flexibility of the potential energy surface and can reproduce \textit{ab initio} energies and forces with high accuracy~\citep{unke2021machine}. Nevertheless, a machine learning force field trained only in a bottom-up manner inherits the limitations of its reference electronic-structure method. 
Recent experimental benchmarks have highlighted this discrepancy, revealing that a potential that accurately fits quantum-mechanical forces is not guaranteed to reproduce experimental observables~\citep{mannan_uniffbench_2026}.
This motivates an increasingly important direction: refining force fields directly against experimental measurements~\citep{frohlking2020toward}.

Recent approaches have demonstrated the feasibility of experimental alignment using time-independent observables. Differentiable Trajectory Reweighting (DiffTRe) enables the optimization of machine learning force fields against static observables without differentiating molecular trajectories~\citep{difftre}. Subsequent studies have combined quantum-mechanical and experimental data to refine machine learning force fields~\cite{rocken2024accurate}, and experimental density alignment has improved the prediction of properties of liquid electrolytes~\cite{gong2025predictive}. However, despite these advances in aligning static experimental properties, directly incorporating dynamical observables, such as diffusion coefficient and viscosity, into force field optimization remains considerably more challenging.

Dynamical observables provide information about molecular motion and relaxation processes that cannot be fully characterized by static properties alone. Unlike static structural and thermodynamic quantities, which can be evaluated from instantaneous configurations sampled at equilibrium, dynamical observables depend on the temporal correlation along molecular trajectories. 
A natural approach is to differentiate these trajectories with respect to the force field parameters. However, molecular dynamics is inherently chaotic and small parameter perturbations can lead to exponentially growing trajectory sensitivities, making long-time gradient estimation numerically unstable~\citep{metz2021}.
Recent methods based on adjoint differentiation\cite{raman} and reversible molecular simulation~\citep{langevin} have enabled the use of dynamical observables in force field optimizations. Nevertheless, these approaches retain the sensitivity of individual trajectories to parameter perturbations and typically require truncation of gradient propagation or correlation functions to control numerical instabilities. Such approximations can compromise gradient accuracy, particularly for observables governed by long-time dynamics.

A conceptually different route is to differentiate the probability of a path rather than the path itself. Such score-function (likelihood-ratio) estimators~\citep{glynn1990,mohamed2019} do not inherit the sensitivity of chaotic dynamics. For Langevin dynamics, the path probability follows from the Girsanov/Onsager–Machlup action and has been used for path reweighting~\citep{donati2017,kieninger2021path} and parametric sensitivity analysis~\citep{warren2012malliavin,warren2012steady}. Bolhuis et al.~\citep{bolhuis2023} used path reweighting to impose rate constants on force field parameters, but only for overdamped dynamics of low-dimensional models with up to two parameters. Whether the path score can drive many-parameter force field optimization against transport coefficients has remained open.

Here we develop PathGrad, a path-probability gradient framework for optimizing force fields against experimental dynamical observables. Building on likelihood-ratio sensitivity analysis, PathGrad evaluates parameter gradients from the probability measure of discretized Langevin trajectories, using thermal noise and force derivatives accumulated during simulation.
Because static observables enter through the same covariance form with the configurational score, dynamical and structural targets can be optimized together.
We first validate the estimator on a two-dimensional toy model against finite differences, and then apply PathGrad to an embedded-atom model (EAM)~\citep{eam} of molten copper and a neuroevolution potential (NEP)~\citep{fan2021neuroevolution} for water.
For molten copper, PathGrad fine-tunes a 12-parameter EAM potential to the experimental diffusion coefficient while preserving the lattice constant and bulk modulus.
For water, an NEP model is optimized against the experimental diffusion coefficient and O--O radial distribution function (RDF). The O--H and H--H RDFs, the shear viscosity, and the rotational relaxation time also improve, although none of them enters the loss. 
Remarkably, alignment at a single temperature also improves the apparent activation energies of these dynamical observables, bringing their temperature dependences into substantially better agreement with experiment.
Moreover, the force error relative to the CCSD(T)-level reference data is decreased, which provides evidence that this experimental alignment improves the fidelity of the potential energy surface beyond the fitted targets.
These results support PathGrad as a general framework for refining force fields directly against experimental observables.

\section{Results}
\label{sec:results}

\subsection{Path-probability gradient (PathGrad).}

We consider a force field $\mathbf F_\theta(\mathbf r) = - \nabla_{\mathbf r} U_\theta(\mathbf r)$ with parameters $\theta$, where $\mathbf r$ denotes the coordinates of all particles in the system.
Static properties like pressure and radial distribution functions depend only on $\mathbf r$ and $\theta$, while dynamical properties like diffusion coefficient depend on the whole trajectory $\Gamma=\{\mathbf r_t\}$. The gradient of an observable $O$ against the parameters can be calculated using
\begin{equation}\label{eq:main}
\nabla_\theta\langle O\rangle_\theta = \langle \nabla_\theta O \rangle_\theta + \langle (O - \langle O \rangle) \nabla_\theta \log p_\theta\rangle_\theta
\end{equation}
where $p_\theta$ is the probability density function. 
For static observables, this reduces to the thermodynamic-fluctuation formula underlying ensemble-reweighting schemes such as DiffTRe~\citep{difftre}. 
For dynamical observables, the most intuitive way is to consider the dynamical observable as a function of the initial positions and the force field parameters. In this case, we need to differentiate the trajectory directly using methods like differentiable simulation~\citep{jaxmd,dmff}. However, along MD trajectories, the variance of this gradient grows exponentially with trajectory length, thereby hindering its practical applicability. 
To overcome this limitation, we propose to treat the whole trajectory instead of the initial positions as random variable. In this case, the first term on the right hand side of Eq.~\eqref{eq:main} vanishes, and now the goal is to derive the probability density for a trajectory. We consider the discretized Langevin dynamics
\begin{align}
\mathbf{r}_{n+1} &= \mathbf{r}_n + \mathbf{v}_n\Delta t \\
\label{velUpdate}
\mathbf{v}_{n+1} &= \mathbf{v}_n
+ \left[\frac{\mathbf{F}_\theta(\mathbf{r}_n)}{M} - \gamma\mathbf{v}_n\right]\Delta t
+ \sqrt{\frac{2\gamma k_B T\Delta t}{M}}\,\xi_n
\end{align}
where $\xi_n \sim \mathcal{N}(0,\mathbf{I})$ are independent standard Gaussian vectors.
For simplicity of presentation we use the Euler-Maruyama scheme here; the numerical implementation uses the A-BO Langevin integrator~\citep{abo1,abo2} detailed in the Supplemental Material.
For a trajectory $\Gamma = \{\mathbf{r}_n,\, \mathbf v_n\}$, the noise at each step can be inferred by inverting Eq.~\eqref{velUpdate}:
\begin{equation}\label{infer}
\xi_n(\Gamma, \theta) = \sqrt{\frac{M\Delta t}{2\gamma k_B T}}\,
\Bigl[\frac{\mathbf{v}_{n+1} - \mathbf{v}_n}{\Delta t}
+ \gamma\mathbf{v}_n - \frac{\mathbf{F}_\theta(\mathbf{r}_n)}{M}\Bigr]
\end{equation}

Since each $\xi_n$ is a standard Gaussian, the path probability with given initial conditions $\mathbf r_0$ and $\mathbf v_0$ is
\begin{equation}\label{action}
p_\theta\!\left(\Gamma |\, \mathbf r_0,\mathbf v_0\right) = \prod_n \exp\Bigl[-\tfrac{1}{2}\xi_n(\Gamma, \theta)^2\Bigr],
\end{equation}

Note that the normalization constant can be ignored since Eq.~\eqref{eq:main} doesn't require the probability density to be normalized. This probability density becomes the Onsager--Machlup action in the limit $\Delta t \to 0$. The total probability is $p_\theta(\Gamma) = p_\theta(\Gamma|\mathbf r_0, \mathbf v_0)\, p_\theta(\mathbf r_0, \mathbf v_0)$ where $p_\theta(\mathbf r_0, \mathbf v_0)$ is the Boltzmann distribution. The gradient of the log probability is
\begin{equation}
\nabla_\theta \log p_\theta(\Gamma) = S_\theta[\Gamma] - \beta \nabla_\theta U(\mathbf r_0)
\end{equation}
where $S_\theta[\Gamma] = \nabla_\theta \log p_\theta\!\left(\Gamma| \mathbf r_0,\mathbf v_0\right)$ is the path score function. From Eq.~\eqref{infer} and Eq.~\eqref{action}, we have
\begin{equation}\label{pathScore}
S_\theta[\Gamma]
= \sqrt{\frac{\Delta t}{2\gamma M k_B T}}\, \sum_n \xi_n \cdot \nabla_\theta\mathbf{F}_\theta(\mathbf{r}_n)
\end{equation}

This path score $S_\theta[\Gamma]$ is computed on the fly accumulating $\xxi_n$ and
$\nabla_\tht\F_\tht(\rr_n)$ stepwise during a standard Langevin simulation. It requires no derivative through the trajectory, thereby avoiding the Lyapunov instability of chaotic dynamics.
The fine-tuning process is as follows: in each iteration, we run multiple independent simulations in parallel. During the simulation, we not only evolve the positions and velocities, but also accumulate the path score function according to Eq.~\eqref{pathScore}. Then we calculate the dynamics observables for each simulation, and use Eq.~\eqref{eq:main} to calculate the gradient.

\begin{figure*}[t]
\centering
\includegraphics{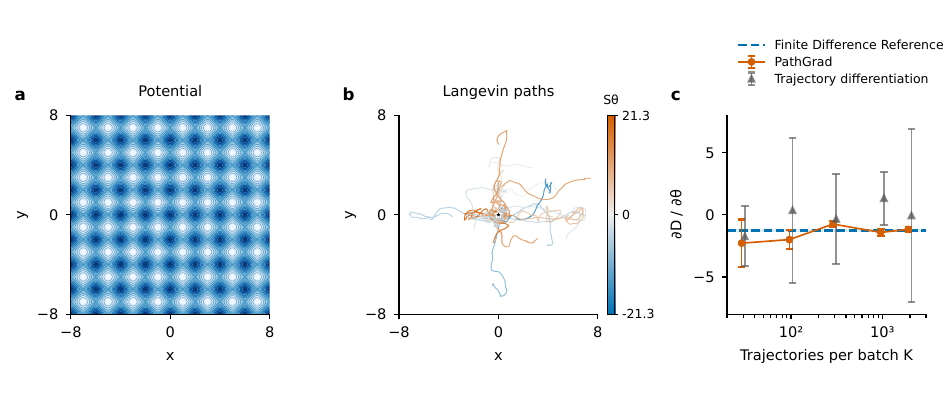}
\caption{\textbf{Path-score differentiation on a two-dimensional toy model.}
(\textbf{a})~Corrugated potential $U(x,y;\tht)=-\tht[\cos(\pi x)+\cos(\pi y)]$ with $\tht=0.5$. Darker regions denote lower potential energy; the barrier between neighboring minima is $2\tht$. 
(\textbf{b})~Twenty-four illustrative Langevin trajectories starting from rest at the origin, colored by their path score $S_\theta[\Gamma_i]$. The gradient is determined by the covariance between the diffusion estimate and the path score. 
(\textbf{c})~Estimates of $\partial D/\partial\theta$ from PathGrad and direct trajectory differentiation as a function of the number of trajectories per batch. Points and error bars denote the mean and sample standard deviation over five independent batches. The blue dashed line and shaded band show the finite-difference reference, $-1.261\pm0.093$, obtained from five batches with $K=2000$.
}
\label{fig:method}
\label{fig:toy}
\end{figure*}

\subsection{Toy model.}

We first validate the path-score estimator on a system for which the gradient can be independently estimated by finite differences. 
A single particle evolves under Langevin dynamics on a two-dimensional corrugated potential $U(x,y;\tht)=-\tht[\cos(\pi x)+\cos(\pi y)]$, where $\tht$ sets the height of the barriers between adjacent minima (Fig.~\ref{fig:toy}a). 
Increasing $\tht$ suppresses thermally activated hopping and lowers the diffusion coefficient $D$, so $\partial D/\partial\theta<0$. We propagate $K$ independent trajectories from $\mathbf{r}_0=(0,0)$ and estimate $D$ from their mean-squared displacement (MSD).

Figure~\ref{fig:toy}b illustrates different trajectories. Each trajectory $\Gamma_i$ is colored by its path score $S_\tht[\Gamma_i]=\partial_\tht\log p_\tht[\Gamma_i]$.
A positive score means that a small increase in $\tht$ increases the relative likelihood of this path, whereas a negative score means that its likelihood decreases. 
Trajectories that diffuse farther tend to have negative scores, whereas less mobile trajectories tend to have positive scores. Their negative covariance $\langle(D_i-\langle D\rangle)\,\partial_\tht\log p_\tht[\Gamma_i]\rangle$ therefore gives $\partial D/\partial\theta<0$. 
The covariance quantifies how a small change in $\tht$ redistributes the probability between high- and low-mobility paths. 

Figure~\ref{fig:toy}c compares the path-score estimator with direct trajectory differentiation as the batch size $K$ increases. 
The path-score estimate converges to the finite-difference reference, and its statistical uncertainty decreases approximately as $1/\sqrt{K}$.
In contrast, direct trajectory differentiation remains highly variable even at $K=2000$, giving $-0.06\pm6.94$, whose spread is much larger than the gradient value itself. 
Near a barrier top, the dynamics contains an unstable direction, so small parameter perturbations can be strongly amplified and can sharply alter the timing or occurrence of a crossing event~\citep{metz2021,suh2022}. Path-score differentiation estimates the gradient from changes in the statistical weights of sampled paths, and therefore avoids propagating this trajectory sensitivity.

\subsection{Fine-tuning an embedded-atom potential for molten copper.}

The embedded-atom model (EAM) of Mishin \emph{et al.}~\citep{eam} was fitted to lattice structure, elastic properties, and \textit{ab initio} data, and accurately describes defects in copper. Here we use a 12-parameter analytic representation fitted to the Mishin potential (Methods), which reproduces the crystalline lattice properties while failing to reproduce the experimental diffusion coefficient of the melt. 
We use it as a first atomistic demonstration, fine-tuning the potential to recover the target diffusion coefficient while preserving its structural fidelity.

Specifically, we fine-tune the potential against three targets simultaneously: the diffusion coefficient $D$, the lattice constant $a_0$, and the bulk modulus $B$. The structural constraints are applied through derivatives of the total energy $U(L)$ of a $4\times4\times4$ FCC supercell with respect to its side length $L$. At the equilibrium supercell side $L_0=4a_0$ the pressure vanishes, giving the bulk modulus as $B=(9L)^{-1}\partial^2 U/\partial L^2|_{L=L_0}$. The combined loss is
\begin{equation}\label{eq:eamloss}
\mathcal{L}
=\lambda_1\bigl(D-D_\mathrm{target}\bigr)^2
+\Bigl[\lambda_2\Bigl(\tfrac{\partial U}{\partial L}\Bigr)^2
+\lambda_3\Bigl(\tfrac{\partial^2 U}{\partial L^2}-9BL\Bigr)^2\Bigr]_{L=L_0}\!,
\end{equation}
where $D_\mathrm{target}$ is the experimental target~\citep{Cu0,Cu1,Cu2,Cu3} and $\lambda_1, \lambda_2, \lambda_3$ are adjustable weights. 

\begin{figure}[t]
\centering
\includegraphics{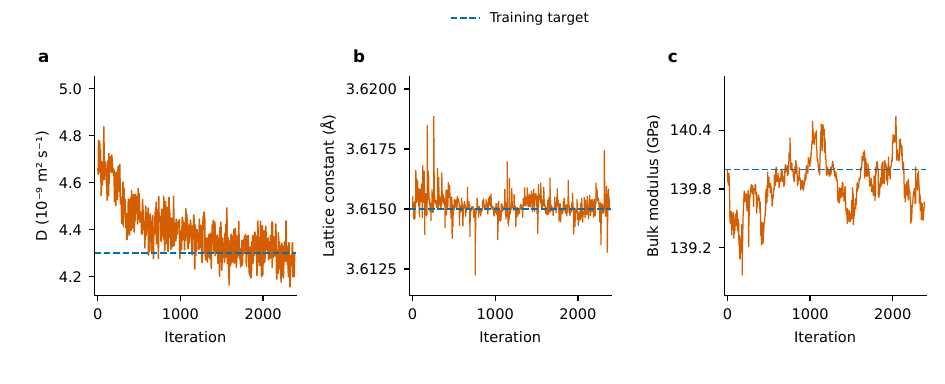}
\caption{\textbf{Fine-tuning the EAM potential for molten copper.}
(\textbf{a})~Langevin diffusion estimate $D$ versus optimization iteration, with the experimental target $D_\mathrm{target}=4.3\times10^{-9}\,\mathrm{m^2\,s^{-1}}$ (dashed). 
(\textbf{b})~Equilibrium lattice constant $a_0$ and (\textbf{c})~bulk modulus $B$ versus optimization iteration, with targets $a_0=3.615\,\text{\AA}$ and $B=140\,\mathrm{GPa}$ (dashed).}
\label{fig:cu}
\end{figure}

The optimization trajectory is shown in Fig.~\ref{fig:cu}. 
The diffusion coefficient starts from a larger value and converges to the target within the first few hundred iterations. The structural constraint is transiently perturbed while $D$ is being corrected and is then gradually restored over subsequent iterations. 

\subsection{Fine-tuning a machine learning force field for water.}

We next apply PathGrad to a machine learning force field, where the parameter space is larger and the potential is more flexible. The neuroevolution potential (NEP)~\citep{fan2021neuroevolution} offers high computational efficiency while retaining fidelity to reference quantum chemistry data. When trained on MB-pol reference data~\citep{babin2013development,babin2014development} that approach the accuracy of CCSD(T), which is the benchmark "gold standard" of molecular quantum chemistry~\citep{bartlett2007coupled}, an NEP model can accurately reproduce the experimental RDF, density, and diffusion coefficient of water~\citep{mbpol}. When trained on density functional theory (DFT) data with the SCAN meta-generalized-gradient-approximation functional~\citep{sun2015scan,chen2017water}, however, the diffusion coefficient is significantly lower than the experimental target (Supplementary Table~S2).
Starting from a publicly available NEP model~\citep{fan2026qnep} trained on the SCAN dataset~\citep{xu2023accurate}, we fine-tune it to recover the experimental diffusion coefficient while simultaneously fitting the O--O radial distribution function (RDF) to prevent structural degradation, using the combined loss
\begin{equation}
\mathcal{L}=(D-D_\mathrm{target})^2
+\lambda_S\sum_i\bigl[g(r_i)-g_0(r_i)\bigr]^2 ,
\end{equation}
where $g(r)$ is computed per frame as a Gaussian-smeared RDF (width $\varepsilon=0.1\,\text{\AA}$) and $g_0(r)$ is the neutron-scattering reference~\citep{neutron} smoothed with the same width. Gaussian smearing provides a smooth, differentiable target and suppresses the high-variance binning noise of a conventional histogram. The RDF gradient follows from Eq.~\eqref{eq:main} using the configurational score $-\beta\nabla_\tht U_\tht(\mathbf{r})$ with $\beta=1/(k_\mathrm{B}T)$. 

\begin{figure*}[t]
\centering
\includegraphics{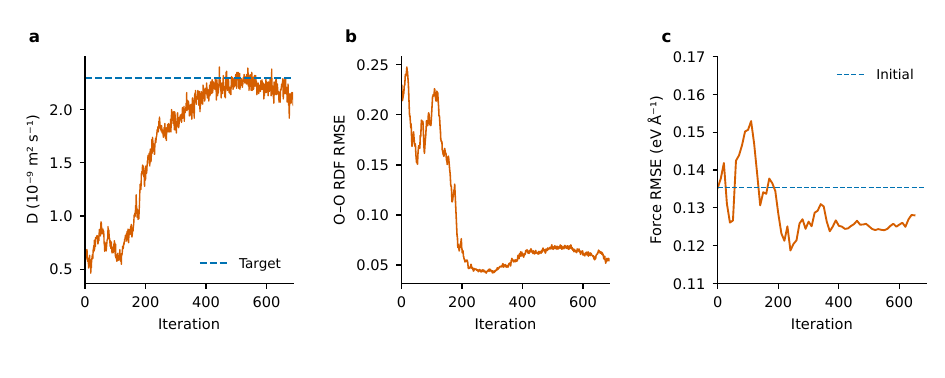}
\caption{\textbf{Fine-tuning the water potential.}
(\textbf{a})~Diffusion during optimization, with its target (dashed).
(\textbf{b})~O--O RDF RMSE during optimization.
(\textbf{c})~Force RMSE against the CCSD(T)-level reference~\citep{mbpol}, evaluated every 10 iterations.
}
\label{fig:water_kinetics}
\end{figure*}

\begin{figure*}[t]
\centering
\includegraphics{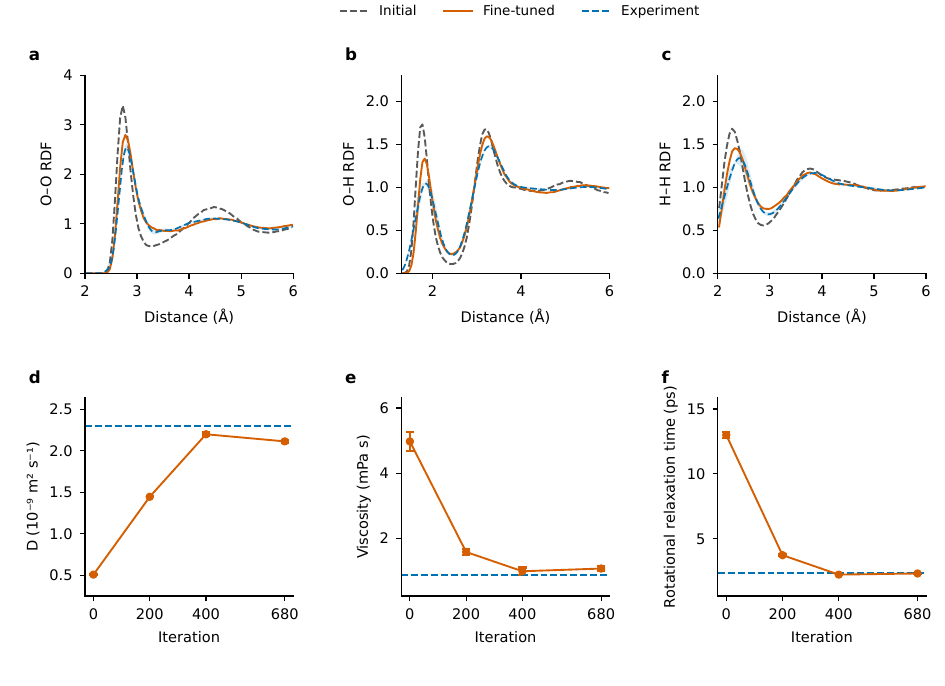}
\caption{\textbf{Water structure and dynamics before and after fine-tuning.}
All results are evaluated at $T=298.15\,\mathrm{K}$ and $\rho=0.99704\,\mathrm{g\,cm^{-3}}$.
(\textbf{a}--\textbf{c})~O--O, O--H and H--H RDFs for the initial and fine-tuned (at iteration 680) models. 
Blue dashed curves are experimental references: O--O from X-ray scattering at $299.5\,\mathrm{K}$~\citep{skinner2014structure}; O--H and H--H from empirical potential structure refinement (EPSR) of scattering data at $298\,\mathrm{K}$~\citep{soper2013rdf}. 
Shading shows the uncertainty or variation reported by the corresponding sources. 
Only intermolecular regions are shown for O--H ($r\geq1.3\,\text{\AA}$) and H--H ($r\geq2.0\,\text{\AA}$).
(\textbf{d}--\textbf{f})~Diffusion coefficient $D$, shear viscosity $\eta$ and H--H rotational relaxation time $\tau_2$ at iterations $0$, $200$, $400$ and $680$. 
Symbols are means over five independent $5\,\mathrm{ns}$ trajectories and error bars are sample standard deviations. 
Some error bars are too small to be visible in the figure. 
Dashed lines are experimental values~\citep{H2O_1,korson1969,jonas1976,kumar2008}.
The optimization loss included only $D$ and the O--O RDF. 
The estimation windows and underlying curves are shown in Supplementary Fig.~S2.}
\label{fig:water_nonaligned}
\label{fig:water_structure}
\end{figure*}

The optimization results are shown in Fig.~\ref{fig:water_kinetics}. 
The diffusion coefficient rises from $0.61\times10^{-9}\,\mathrm{m^2\,s^{-1}}$ and reaches the training target of $2.3\times10^{-9}\,\mathrm{m^2\,s^{-1}}$ after about 450 iterations (Fig.~\ref{fig:water_kinetics}a).
The root-mean-square error (RMSE) of the RDF also decreases from $0.230$ to $0.055$ (Fig.~\ref{fig:water_kinetics}b). Correspondingly, the final O--O RDF visibly approaches the X-ray-scattering reference~\citep{skinner2014structure} (Fig.~\ref{fig:water_structure}a), and the O--H and H--H RDFs, which are not constrained in the loss, also approach their experimental references (Fig.~\ref{fig:water_structure}b,c). 
We also calculate the force RMSE relative to the CCSD(T)-level MB-pol reference dataset~\citep{mbpol} every 10 iterations (Fig.~\ref{fig:water_kinetics}c).
The force RMSE is observed to decrease by $7.5\%$ over the first 450 iterations, even though the CCSD(T) forces were not included in the loss. This suggests that correcting the dynamics does not come at the expense of force field fidelity; on the contrary, the fine-tuned model moves closer to the high-level reference in force space as well.

We then evaluated the shear viscosity $\eta$ and the second-order rotational relaxation time $\tau_2$ of the intramolecular H--H vector against experimental measurements to determine whether fine-tuning improved the description of observables not included in the loss function. As shown in Fig.~\ref{fig:water_nonaligned}d--f, the initial model exhibits slow diffusion and reorientation and excessive viscosity; all properties improve after fine-tuning. 

\begin{figure*}[t]
\centering
\includegraphics{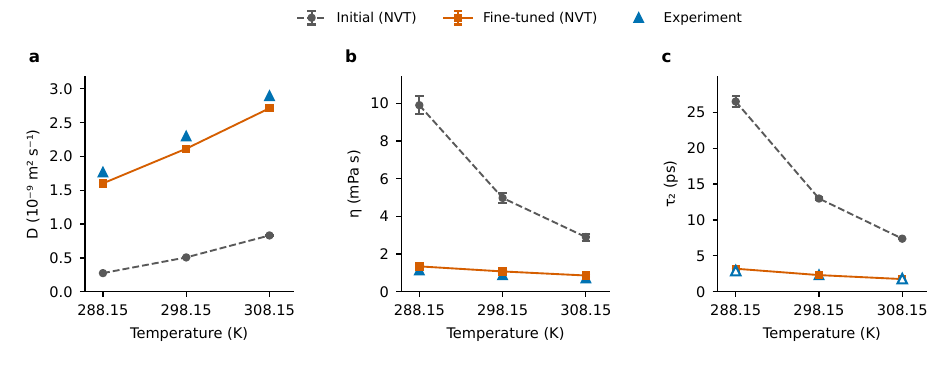}
\caption{\textbf{Temperature transferability after fine-tuning at 300 K.}
(\textbf{a})~Diffusion coefficient $D$, (\textbf{b})~shear viscosity $\eta$ and (\textbf{c})~rotational relaxation time $\tau_2$ at $288.15$, $298.15$ and $308.15\,\mathrm{K}$, before (initial, grey) and after (fine-tuned, orange) optimization at 300 K (Methods). 
Simulations at each temperature were performed in the NVT ensemble using the corresponding experimental density~\citep{kell1975}. 
Symbols and error bars denote the mean and sample standard deviation, respectively, over five independent trajectories.
Some error bars are too small to be visible in the figure. 
Experimental references~\citep{H2O_1,korson1969,jonas1976,kumar2008} are shown as triangles. 
Open triangles mark the $\tau_2$ references at $288.15$ and $308.15\,\mathrm{K}$, digitized from a fitted experimental curve~\citep{krynicki1966} (see Supplementary Tab.~S3).}
\label{fig:water_temperature}
\end{figure*}

Because the model was fine-tuned using trajectories at a single temperature of 300 K, we also evaluated its performance at other temperatures. Without retraining, we evaluated $D$, $\eta$, and $\tau_2$ at $288.15$, $298.15$ and $308.15\,\mathrm{K}$, each at the experimental density at that temperature~\citep{kell1975} (Fig.~\ref{fig:water_temperature}). Before fine-tuning, the discrepancy with experiment grows on cooling; after fine-tuning, the model stays close to the measured values over the whole $20\,\mathrm{K}$ window, although none of these temperatures entered the training.
Notably, fine-tuning also brings the apparent activation energies of these dynamical observables substantially closer to experimental values (Supplementary Fig.~S3 and Table~S4). Thus, single-temperature experimental alignment improves not only the magnitudes of molecular transport and relaxation but also their temperature dependences, suggesting a physically meaningful refinement of the underlying potential energy surface.

\begin{figure*}[t]
\centering
\includegraphics[width=\textwidth]{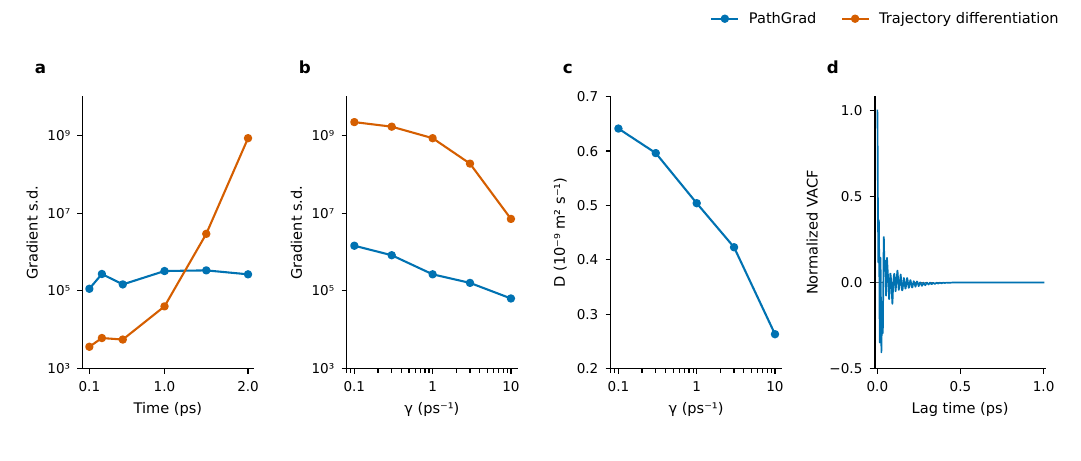}
\caption{\textbf{Gradient standard deviation and dynamics of the initial NEP water model.}
(\textbf{a})~Gradient standard deviation versus trajectory length at $\gamma=1\,\mathrm{ps^{-1}}$. The results for longer trajectories are presented in Supplementary Fig.~S1. 
(\textbf{b})~Gradient standard deviation versus $\gamma$ for $2\,\mathrm{ps}$ trajectories.
Both methods estimate the gradient of the diffusion observable with respect to the potential parameters.
(\textbf{c})~Diffusion coefficient $D$ versus $\gamma$. Each point is obtained from a single $5\,\mathrm{ns}$ trajectory using oxygen MSD fit.
(\textbf{d})~All-atom velocity autocorrelation function (VACF), normalized by its ensemble mean at zero lag, at $\gamma=1\,\mathrm{ps^{-1}}$.
}
\label{fig:water_variance}
\end{figure*}

\section{Discussion}
\label{sec:discussion}

Differentiating path probabilities rather than the trajectories themselves avoids the exponential growth of gradient variance with trajectory length that arises in chaotic molecular dynamics. 
For the NEP water model, the variance of direct trajectory differentiation over a $2\,\mathrm{ps}$ trajectory is already several orders of magnitude larger than the PathGrad variance (Fig. 6a) and the gradient overflows numerically from the very first optimization step.
The variance of the path score, by contrast, grows only linearly (see Supplemental Material), as expected for likelihood-ratio estimators~\citep{glynn1990,arampatzis2016efficient}. 
At the same time, the statistical error of a diffusion estimate averaged over time origins decreases with $t$. These two trends largely compensate, so the variance of the gradient estimator depends only weakly on the trajectory length, as shown in Supplementary Fig.~S1.

An important practical consideration is the choice of the Langevin friction coefficient $\gamma$, which sets a trade-off between gradient noise and dynamical fidelity. At fixed trajectory length, the variance of the path score scales as $1/\gamma$ (Fig. 6b), so smaller friction requires more independent trajectories for training. Larger friction gives less noisy gradients, but perturbs transport more strongly (Fig. 6c). 
Smaller values of $\gamma$ interfere less with the intrinsic dynamics and yield results closer to those obtained in the microcanonical (NVE) ensemble. 
The value used here, $\gamma=1\ \mathrm{ps}^{-1}$, is a common choice for Langevin thermostats and was also used for reversible simulation~\cite{langevin}. 

The length of each independent trajectory is another key parameter. 
Because the PathGrad variance remains almost unchanged with trajectory length (Fig. 6a and Supplementary Fig.~S1), the trajectory length can be chosen by physical rather than numerical considerations: trajectories should be sufficiently long for the target observable to enter the relevant long-time dynamical regime. 
We selected 1 ps for copper and 2 ps for water according to the decay of the velocity autocorrelation function (VACF) (Fig. 6d), which is sufficient for the diffusion coefficient. 
Observables with slower relaxation, such as shear viscosity or reorientation in supercooled liquids, require longer windows, and for these targets the weak length dependence of the PathGrad variance is a particular advantage. 
When additional computational resources were available, we increased the number of independent trajectories rather than their length, since the gradient variance decreases as $1/K$.

Machine-learning force fields are trained on ever larger electronic-structure datasets, yet their accuracy remains bounded by the reference method, usually density functional theory. Learning from experiment is the most direct route beyond this ceiling. 
The experimental observables that MD simulation predicts fall broadly into structural quantities and dynamical quantities. 
Configurational reweighting~\citep{difftre} already provides efficient and general gradients for structural observables, whereas differentiating dynamical observables remains a challenge~\citep{bolhuis2023,raman,langevin}. 
PathGrad differentiates both structural and dynamical observables through a single, untruncated gradient estimator accumulated within a standard Langevin simulation. In principle, it applies to any observable that can be expressed as an average over finite-length Langevin paths.
The water results suggest what this enables. In particular, the force error relative to CCSD(T)-level reference data decreased by about 7.5\%, despite the absence of quantum-chemical data from the loss function. 
PathGrad is thus a unified gradient-based route to using both structural and dynamical experimental data as direct learning targets for machine-learning force fields, removing gradient instability that limits the time window of dynamical training targets.

\FloatBarrier
\clearpage

\section*{Methods}
\label{sec:methods}

\subsection{Simulation and fine-tuning protocols}

All simulations use the Langevin A--BO discretization and the system-specific parameters are collected in Table~S5. 

\emph{Toy model.} In dimensionless units, we set $M=T=1$, $\gamma=0.5$, $\Delta t=0.005$ and $\tht=0.5$. Each trajectory starts from rest at the potential minimum $\rr_0=(0,0)$ and is integrated for $2000$ steps. Central finite differences use perturbations $\tht\pm0.02$ with common random numbers. The path-score covariance centers both $D_k$ and $\score[\Gamma_k]$ within each batch. Error bars report the sample standard deviation across the five independent batch estimates.

\emph{Copper.} We use a 12-parameter analytic representation fitted to the EAM potential of Mishin \emph{et al.}~\citep{eam}, comprising an embedding function $F(\bar\rho)=F_C[\exp(-F_k\bar\rho)-1]$ (two parameters), a pair potential $\psi(r)=\sum_{i=1}^{4}C_i\exp(-k_ir)$ (eight parameters), and a host electron-density function $\rho(r)=C_\rho\exp(-k_\rho r)$ (two parameters), where $\bar\rho$ is the sum of neighbor electron densities. We simulate $256$ copper atoms at $T=1500\,\mathrm{K}$ and mass density $\rho=7.94\,\mathrm{g\,cm^{-3}}$ with $\gamma=1\,\mathrm{ps^{-1}}$ and $t=1\,\mathrm{ps}$, using the combined loss in Eq.~\eqref{eq:eamloss}. The structural constraint is evaluated from the total energy $U(L)$ of a $4\times4\times4$ FCC supercell ($256$ atoms) with side length $L$ evaluated at $L_0=4a_0=14.46\,\text{\AA}$, corresponding to the experimental lattice constant $a_0=3.615\,\text{\AA}$ and bulk modulus $B=140\,\mathrm{GPa}$~\citep{CuLat,CRC}. 
In each optimization step, the dynamical gradient is projected orthogonal to the structural gradient before updating the parameters.

\emph{Water.} Starting from a publicly available NEP model~\citep{fan2026qnep} trained on DFT data generated with the SCAN functional~\citep{xu2023accurate}, we simulate $512$ water molecules ($1536$ atoms) at $\rho=0.994\,\mathrm{g\,cm^{-3}}$ with $\gamma=1\,\mathrm{ps^{-1}}$ and trajectory length $t=2\,\mathrm{ps}$.
The Langevin thermostat target is set to $T=300\,\mathrm{K}$ with initial velocities drawn from the Boltzmann distribution, resulting in a recorded kinetic temperature during training between $297$ and $301\,\mathrm{K}$ (mean $299\,\mathrm{K}$). 
The structural target $g_0(r)$ is the neutron-scattering O--O RDF~\citep{neutron}, Gaussian-smeared with width $\varepsilon=0.1\,\text{\AA}$, and $g(r)$ is evaluated per frame with the
same kernel. 

\subsection{Transport-coefficient analysis}

The diffusion coefficient is defined by the Einstein relation:
\begin{equation}
D=\lim_{t\to\infty}\frac{1}{6t}
\left\langle|\rr(t_0+t)-\rr(t_0)|^2\right\rangle,
\end{equation}
where the angle brackets denote an average over both time origins $t_0$ and atoms. In practice, we extract $D$ by fitting the linear regime of the mean-squared displacement (MSD).

The shear viscosity of water is obtained from the Green--Kubo relation for the autocorrelation of the symmetric traceless stress tensor $\Pi_{\alpha\beta}=(\sigma_{\alpha\beta}+\sigma_{\beta\alpha})/2-\delta_{\alpha\beta}\operatorname{tr}(\sigma)/3$. 
Because the liquid water is isotropic, all components of $\Pi$ can be combined to improve statistics~\citep{gonzalez2010}, giving the running integral
\[
\eta(t)=\frac{V}{10k_BT}\int_0^t
\sum_{\alpha,\beta}\langle
\Pi_{\alpha\beta}(t_0)\Pi_{\alpha\beta}(t_0+s)\rangle_{t_0}\,\mathrm ds,
\]
where $V$ is the volume and the sum runs over all nine Cartesian component pairs $(\alpha,\beta)$, including both orders of off-diagonal indices, which accounts for the prefactor $1/10$. In practice we accumulate the running integral over lag time and
identify $\eta$ with its plateau mean over $25$--$75\,\mathrm{ps}$.

Molecular reorientation is characterized by the second-order Legendre time-correlation function of the unit intramolecular H--H vector $\hat u$:
\begin{equation}
\tau_2=\int_0^\infty C_2^{\mathrm{HH}}(t)\,\mathrm{d}t,
\qquad
C_2^{\mathrm{HH}}(t)=\bigl\langle P_2\bigl(\hat u(t_0)\cdot\hat u(t_0+t)\bigr)\bigr\rangle ,
\end{equation}
where $P_2(x)=\tfrac12(3x^2-1)$ is the second Legendre polynomial, and the average is taken over all water molecules and time origins. In the extreme-narrowing regime applicable to liquid water near room temperature, $\tau_2$ is the rotational correlation time governing intramolecular $^1\mathrm{H}$ spin--lattice relaxation~\citep{bpp1948,jonas1976,kumar2008}. We evaluate the integral numerically using the trapezoidal rule from $t=0$ out to a lag beyond which the integral no longer changes. At $298.15\,\mathrm{K}$, the recorded windows are $149.5$, $69.5$, $39.5$ and $29.5\,\mathrm{ps}$ for iterations $0$, $200$, $400$ and $680$, respectively; at $288.15\,\mathrm{K}$ the initial and fine-tuned windows are $349.5$ and $49.5\,\mathrm{ps}$, and at $308.15\,\mathrm{K}$ they are $149.5$ and $29.5\,\mathrm{ps}$.

\noindent{\bfseries References}\setlength{\parskip}{12pt}%
\bibliography{references}

\begin{addendum}
 \item[Acknowledgements] This work was supported by the National Natural
       Science Foundation of China (Grant No. 12404291) and the New
       Generation Artificial Intelligence National Science and Technology
       Major Project (2025ZD0121802). Z.W. and L.S. did this work during their
       internship at Shanghai Artificial Intelligence Laboratory.
 \item[Author contributions] M.S. and R.S. conceived the idea and designed the study.
       L.S. and Z.W. developed the path-score formalism and implemented the code.
       Y.Z. performed the transport-coefficient simulations and analysis. 
       X.G., R.S., and M.S. contributed to the analysis and provided scientific insights. 
       L.S. wrote the first draft. 
       All authors reviewed and approved the final manuscript.
 \item[Competing interests] The authors declare no competing interests.
 \item[Correspondence] Correspondence and requests for materials should be
       addressed to R.S.~(ruishi@zju.edu.cn) or M.S.~(sumao@pjlab.org.cn).
\end{addendum}

\end{document}

\maketitle
\tableofcontents
\listoffigures
\listoftables

\clearpage

\section{Theory}\label{sm:sec:theory}

\subsection{Gradient of an ensemble average}\label{sm:sec:gradexp}

We consider a force field with parameters $\theta$ and denote a configuration of the system by $\mathbf r$, with probability density $p_\theta(\mathbf r)$.
For an observable $O$ that depends on $\mathbf r$ and possibly on $\theta$, its ensemble average is
\begin{equation}\label{sm:avg}
    \langle O \rangle_\theta =
    \frac{\int O(\mathbf r,\theta)\,p_\theta(\mathbf r)\,\mathrm{d}\mathbf r}
         {\int p_\theta(\mathbf r)\,\mathrm{d}\mathbf r}.
\end{equation}
Its parameter gradient is obtained from the quotient rule,
\begin{equation}\label{sm:main}
\begin{aligned}
\nabla_\theta\langle O \rangle_\theta &=
\frac{\int\left(\nabla_\theta O \,p_\theta + O\,\nabla_\theta p_\theta\right)\,\mathrm{d}\mathbf r}{\int p_\theta\, \mathrm{d} \mathbf r}
-\frac{\int \nabla_\theta p_\theta \,\mathrm{d}\mathbf r \int O \,p_\theta \,\mathrm{d}\mathbf r}
    {\left(\int p_\theta \,\mathrm{d}\mathbf r\right)^2} \\
&= \frac{\int \nabla_\theta O \,p_\theta \,\mathrm{d}\mathbf r}{\int p_\theta \, \mathrm{d}\mathbf r}
+ \frac{\int O \,\nabla_\theta \log p_\theta \, p_\theta \,\mathrm{d}\mathbf r}{\int p_\theta \,\mathrm{d}\mathbf r}
-\frac{\int\nabla_\theta \log p_\theta\,p_\theta \,\mathrm{d}\mathbf r}{\int p_\theta\,\mathrm{d}\mathbf r}
\frac{\int O \,p_\theta \,\mathrm{d}\mathbf r}{\int p_\theta\,\mathrm{d}\mathbf r} \\
&= \langle \nabla_\theta O\rangle_\theta
+ \langle O \,\nabla_\theta \log p_\theta\rangle_\theta
- \langle \nabla_\theta \log p_\theta \rangle_\theta
\langle O\rangle_\theta \\
&= \langle \nabla_\theta O\rangle_\theta
+ \big\langle \bigl(O - \langle O \rangle_\theta\bigr)\, \nabla_\theta \log p_\theta \big\rangle_\theta .
\end{aligned}
\end{equation}
This is the score-function or likelihood-ratio gradient estimator used in stochastic simulation, operations research, and reinforcement learning~\cite{glynn1990,williams1992,kleijnen1996,fu2006,mohamed2019}: the score $\nabla_\theta\log p_\theta$ reweights observable fluctuations by the parameter sensitivity of the sampling distribution.
Because the right-hand side of Eq.~\eqref{sm:main} involves only $\nabla_\theta\log p_\theta$, and any additive constant in $\log p_\theta$ drops out of the covariance, we may work with unnormalized densities throughout.
In the following we apply Eq.~\eqref{sm:main} to two sampling distributions: the Boltzmann distribution of single configurations, which yields the configurational score used for structural observables such as the radial distribution function (Sec.~\ref{sm:sec:rdf}), and the probability distribution over Langevin trajectories, which yields the path score used for dynamical observables such as the diffusion coefficient (Sec.~\ref{sm:sec:diffusion}).

\subsection{Configurational score and structural observables}\label{sm:sec:rdf}

For a static observable $Q(\mathbf r)$, such as the radial distribution function (RDF), the sampling distribution is the Boltzmann distribution $p_\theta(\mathbf r)\propto\exp[-\beta U_\theta(\mathbf r)]$ with $\beta=1/(k_BT)$, whose score is
\begin{equation}
\nabla_\theta\log p_\theta(\mathbf r) = -\beta\,\nabla_\theta U_\theta(\mathbf r) + \text{const},
\end{equation}
where the configuration-independent constant (the gradient of the log partition function) drops out of the covariance in Eq.~\eqref{sm:main}.
The gradient of a static ensemble average therefore reduces to the configurational form
\begin{equation}\label{sm:static}
\nabla_\theta\langle Q\rangle_\theta
=
\langle\nabla_\theta Q\rangle_\theta
-\beta\,\big\langle\bigl(Q-\langle Q\rangle_\theta\bigr)\,\nabla_\theta U_\theta\big\rangle_\theta ,
\end{equation}
which is the estimator underlying the DiffTRe method~\cite{difftre}.

The RDF $g(r)$ is evaluated per configuration as a Gaussian-smeared histogram:
\begin{equation}\label{sm:rdf}
g(r_i) = \frac{1}{4\pi r_i^2\rho N}\,
\sum_{a=1}^N\sum_{b\neq a}
\frac{1}{\sqrt{2\pi\varepsilon^2}}\,
\exp\!\Bigl[-\frac{(r_i - |\mathbf{r}_a - \mathbf{r}_b|)^2}{2\varepsilon^2}\Bigr],
\end{equation}
where $\rho$ is the number density and $\varepsilon$ is the smearing width (we use $\varepsilon = 0.1\,\text{\AA}$).
The smearing provides a smooth RDF estimate and reduces binning noise. A conventional histogram would introduce discontinuous bin counts and high-variance shot noise.

\subsection{Path score and dynamical observables}\label{sm:sec:diffusion}

Dynamical observables such as the diffusion coefficient depend on the whole trajectory rather than on a single configuration.
We therefore take the trajectory itself as the random variable in Eq.~\eqref{sm:main}.
Below we first discuss the general structure of the path probability and its score, including the treatment of the initial condition (Sec.~\ref{sm:sec:pathprob}), and then evaluate the score explicitly for the A-BO Langevin scheme used in our numerical implementation (Sec.~\ref{sm:sec:lvv}).

\subsubsection{Path probability, score function, and initial conditions}\label{sm:sec:pathprob}

We denote the mass of particle $i$ by $M_i$ and write the multi-particle Langevin dynamics as
\begin{equation}\label{sm:langevin}
\begin{aligned}
\mathrm d \mathbf r_i &= \mathbf v_i \,\mathrm d t,
\\
\mathrm d \mathbf v_i &= \frac{\mathbf F_i(\mathbf r)}{M_i} \,\mathrm{d}t - \gamma \mathbf v_i \, \mathrm{d}t + \sqrt{\frac{2 \gamma\,k_BT}{M_i}} \, \mathrm{d} \mathbf W_i,
\end{aligned}
\end{equation}
where $\mathbf F_i(\mathbf r)=-\nabla_{\mathbf r_i}U_\theta(\mathbf r)$ and $\mathrm d \mathbf W_i$ is a Wiener process.

A numerical integrator advances the phase point $\mathbf z^{(n)}=(\mathbf r^{(n)},\mathbf v^{(n)})$ to $\mathbf z^{(n+1)}$ over a time step $\Delta t$ by drawing Gaussian noise $\boldsymbol\xi_i^{(n)}\sim\mathcal N(0,\mathbf I)$.
For the integrators considered here, the noise can be inferred a posteriori from the stored positions and velocities, and the map from $\boldsymbol\xi_i^{(n)}$ to the updated phase point has a parameter-independent Jacobian, as shown explicitly in Secs.~\ref{sm:sec:lvv}.
The probability of a trajectory $\Gamma$ conditioned on the initial phase point $\mathbf z_0=(\mathbf r_0,\mathbf v_0)$ is therefore, up to a parameter-independent factor,
\begin{equation}\label{eq:prob0}
p_\theta(\Gamma \mid \mathbf z_0) \propto \prod_{i, n} \exp\Big(-\frac{\big|\boldsymbol\xi_i^{(n)}\big|^2}{2}\Big).
\end{equation}
Because Eq.~\eqref{sm:main} involves only $\nabla_\theta\log p_\theta$, the parameter-independent Jacobian factor and the overall normalization of the density drop out and need not be computed.

The full path probability factorizes as
\begin{equation}\label{eq:factor}
p_\theta(\Gamma) = p_\theta(\Gamma \mid \mathbf z_0)\; p_\theta(\mathbf z_0).
\end{equation}
We define the path score
\begin{equation}\label{sm:scoregeneric}
S_\theta[\Gamma] \equiv \nabla_\theta \log p_\theta(\Gamma \mid \mathbf z_0)
= -\sum_{i, n} \boldsymbol\xi_i^{(n)} \cdot \nabla_\theta \boldsymbol\xi_i^{(n)},
\end{equation}
whose explicit form depends on the integrator through the inferred noise and is derived in Secs.~\ref{sm:sec:lvv}.
\cite{donati2017,kieninger2023,bolhuis2023}.
The score thus records the accumulated mismatch between the noise required to generate the observed path under the original and perturbed force fields.

It remains to specify $p_\theta(\mathbf z_0)$, i.e., how the initial condition is treated.
Two situations arise.
If the initial phase point is held fixed, $p_\theta(\mathbf z_0)$ is parameter-independent and $\nabla_\theta\log p_\theta(\Gamma)=S_\theta[\Gamma]$.
If the initial condition is instead sampled from the equilibrium distribution of the force field,
\begin{equation}
p_\theta(\mathbf z_0) \propto \exp\Big[-\beta\Big(U_\theta(\mathbf r_0)+\sum_i\tfrac{1}{2}M_i|\mathbf v_{i,0}|^2\Big)\Big],
\end{equation}
the velocities follow the parameter-independent Maxwell--Boltzmann distribution, while the positions carry the configurational score of Sec.~\ref{sm:sec:rdf}, giving
\begin{equation}\label{sm:totalscore}
\nabla_\theta \log p_\theta(\Gamma) = S_\theta[\Gamma] - \beta \nabla_\theta U_\theta(\mathbf r_0)
\end{equation}
up to an additive constant.
In the examples presented in this work, the initial configurations are drawn from the equilibrium distribution of the current force field, and both contributions are retained in practice: the diffusion gradient is evaluated with the total score \eqref{sm:totalscore}, so each trajectory contributes its accumulated path score $S_\theta[\Gamma]$ together with the configurational score $-\beta\,\nabla_\theta U_\theta(\mathbf r_0)$ of its initial configuration.

For a dynamical observable $O[\Gamma]$ with no explicit parameter dependence, the first term of Eq.~\eqref{sm:main} vanishes and the gradient is the covariance of the observable with the total score \eqref{sm:totalscore}.
With $K$ independent trajectories $\{\Gamma_k\}_{k=1}^K$ started from initial positions $\mathbf r_{k,0}$, we estimate
\begin{equation}\label{sm:estimator}
\bar{O} = \frac{1}{K}\sum_{k=1}^K O[\Gamma_k],
\qquad
\nabla_\theta\langle O\rangle_\theta \approx
\frac{1}{K-1}\sum_{k=1}^K \bigl(O[\Gamma_k] - \bar{O}\bigr)\,\Bigl(S_\theta[\Gamma_k]-\beta\,\nabla_\theta U_\theta(\mathbf r_{k,0})\Bigr).
\end{equation}
For a squared-error loss $\mathcal{L} = (\langle O\rangle_\theta - O_\mathrm{target})^2$,
\begin{equation}\label{sm:lossgrad}
\begin{aligned}
\nabla_\theta\mathcal{L}
&= 2(\langle O\rangle_\theta - O_\mathrm{target})\,\nabla_\theta\langle O\rangle_\theta \\
&\approx 2(\bar{O} - O_\mathrm{target})\,
\frac{1}{K-1}\sum_{k=1}^K \bigl(O[\Gamma_k] - \bar{O}\bigr)\,\Bigl(S_\theta[\Gamma_k]-\beta\,\nabla_\theta U_\theta(\mathbf r_{k,0})\Bigr).
\end{aligned}
\end{equation}

\subsubsection{A-BO Langevin splitting scheme}\label{sm:sec:lvv}

The numerical experiments in this work use the A-BO Langevin discretization~\cite{abo1,abo2} of Eq.~\eqref{sm:langevin}.
For particle $i$ one step advancing $(\mathbf r_i^{(n)},\mathbf v_i^{(n)})$ to $(\mathbf r_i^{(n+1)},\mathbf v_i^{(n+1)})$ reads
\begin{align}
\mathbf r_i^{(n+1)} &= \mathbf r_i^{(n)} + \mathbf v_i^{(n)}\,\Delta t, \label{eq:lvv-r}\\
\mathbf v_i^{(n+1)} &= c_0\,\mathbf v_i^{(n)} + c_{1,i}\,\boldsymbol\xi_i^{(n)} + c_{2,i}\,\mathbf F_{i,\theta}(\mathbf r^{(n+1)}), \label{eq:lvv-v}
\end{align}
with
\begin{equation}\label{eq:lvv-coefs}
c_0 = e^{-\gamma\Delta t},\qquad
c_{1,i} = \sqrt{\frac{(1-c_0^2)\,k_BT}{M_i}},\qquad
c_{2,i} = \frac{1-c_0}{\gamma M_i},
\end{equation}
and $\boldsymbol\xi_i^{(n)}\sim\mathcal N(0,\mathbf I)$.
This scheme applies a full position-drift step with the current velocity and then updates the velocity with the exact Ornstein--Uhlenbeck propagator for a constant force evaluated at the new position~
\cite{paterlini1998,melchionna2007}.

The Gaussian noise can be inferred from the stored positions and velocities,
\begin{equation}\label{eq:lvv-xi}
\boldsymbol\xi_i^{(n)}
=
\frac{1}{c_{1,i}}
\left[
\mathbf v_i^{(n+1)}
-
c_0\,\mathbf v_i^{(n)}
-
c_{2,i}\,\mathbf F_{i,\theta}(\mathbf r^{(n+1)})
\right].
\end{equation}
The map from the noise $\boldsymbol\xi_i^{(n)}$ to the new velocity is linear with determinant $c_{1,i}^{d}$, where $d$ is the spatial dimension, while the position update is deterministic given the old velocity; the Jacobian relating $p_\theta(\Gamma\mid\mathbf z_0)$ to the Gaussian noise distribution is therefore parameter-independent, and Eq.~\eqref{eq:prob0} holds.
Substituting \eqref{eq:lvv-xi} into \eqref{eq:prob0} gives the discrete Onsager--Machlup action~
\cite{om1,om2}
\begin{equation}\label{sm:omAction}
p_\theta(\Gamma \mid \mathbf z_0) \propto
\exp \left[-\sum_i\frac{1}{2c_{1,i}^2}\sum_n\left(
\mathbf v_i^{(n+1)}
-
c_0\,\mathbf v_i^{(n)}
-
c_{2,i}\,\mathbf F_{i,\theta}(\mathbf r^{(n+1)})
\right)^2\right].
\end{equation}

Differentiating the logarithm gives the path score
\begin{equation}\label{sm:pathScoreExplicit}
\begin{aligned}
S_\theta[\Gamma]
&= \nabla_\theta \Big( - \frac{1}{2}\sum_{i, n} \big|\boldsymbol\xi_i^{(n)}\big|^2 \Big) \\
&= -\sum_{i, n} \boldsymbol\xi_i^{(n)} \cdot \nabla_\theta \boldsymbol\xi_i^{(n)} \\
&= \sum_{i, n} \frac{c_{2,i}}{c_{1,i}}\,
\boldsymbol\xi_i^{(n)} \cdot \nabla_\theta \mathbf F_{i,\theta}(\mathbf r^{(n+1)}) \\
&= \sum_{i, n}
\frac{1}{\gamma}\,\,
\sqrt\frac{1 - e^{-\gamma\Delta t}}{(1 + e^{-\gamma \Delta t})\, k_BT M_i}\,\,\,
\boldsymbol\xi_i^{(n)} \cdot \nabla_\theta \mathbf F_{i,\theta}(\mathbf r^{(n+1)}).
\end{aligned}
\end{equation}
In the small-time-step limit, $(1-e^{-\gamma\Delta t})/(1+e^{-\gamma\Delta t})\to\gamma\Delta t/2$ and the prefactor reduces to $\sqrt{\Delta t/(2\gamma M_i k_BT)}$, recovering the Euler--Maruyama form of the score presented in the main text.

\subsection{Variance}

The primary difficulty in differentiating dynamical observables is the growth of the gradient variance with the trajectory length. In direct differentiation through the equations of motion, derivatives are propagated recursively along the trajectory. In a chaotic many-body system, small changes in the force field may be rapidly amplified, causing the resulting trajectory derivatives to explode with simulation time. 

The path-score method avoids propagating these trajectory derivatives. To examine its variance, we consider the small-time-step limit form of Eq.~\ref{sm:pathScoreExplicit}. The contribution from time step $n$ is
\begin{equation}
s_n=
\sum_i
\sqrt{\frac{\Delta t}
{2\gamma M_i k_{\mathrm B}T}}\,
\xi_i^{(n)}
\cdot
\nabla_\theta F_{i,\theta}\!\left(r^{(n+1)}\right),
\qquad
S_\theta=\sum_{n=0}^{N_t-1}s_n .
\end{equation}

The ordering of the A-BO integrator is important here. The position $r^{(n+1)}$, and hence the force derivative evaluated at that position, is determined before the current noise $\xi^{(n)}$ is applied to update the velocity. Therefore, for a given trajectory history, the sign and direction of each score increment are determined by a newly drawn zero-mean Gaussian noise. Positive and negative contributions cancel on average, giving $\langle s_n\rangle=0$.

The same argument applies to correlations between different time steps. For $n<m$, the force derivative at step $m$ may depend on the earlier trajectory and therefore on the earlier noises, but it is independent of the newly drawn noise $\xi^{(m)}$. Averaging over this new noise makes the cross correlation between $s_n$ and $s_m$ vanish. The accumulated score therefore behaves as a random walk with configuration-dependent step sizes, rather than as an iteratively amplified trajectory derivative. 

As a result, the score variance is the sum of the variances of its individual increments. Averaging over the unit Gaussian noise converts the squared noise projection into the squared norm of the force derivative, giving
\begin{equation}\label{sm:var_s}
\mathrm{Var}(S_\theta)
\simeq
\sum_{n=0}^{N_t-1}\sum_i
\frac{\Delta t}
{2\gamma M_i k_{\mathrm B}T}
\left\langle
\left|
\nabla_\theta F_{i,\theta}(r^{(n+1)})
\right|^2
\right\rangle .
\end{equation}

After equilibration, the average force-gradient norm is approximately stationary. Since $t=N_t\Delta t$, Eq.~\ref{sm:var_s} shows that
\begin{equation}
    \mathrm{Var}(S_\theta)\propto t.
\end{equation}

Thus, the path score has random-walk-like fluctuations whose variance grows linearly rather than exponentially with the trajectory length. The equilibrium initial-condition contribution in Eq.~\ref{sm:totalscore} adds a time-independent term and does not change this long-time scaling.

The gradient estimator involves the product of the score and the observable, $\delta O_tS_\theta$. 
We assume that the observable is evaluated by averaging over particles and time origins, and that the trajectory length substantially exceeds the relevant correlation time, as is standard practice in molecular dynamics simulations.
Under this standard long-trajectory approximation, the number of effectively independent samples is proportional to $t$, and the variance of the observable decreases as
\begin{equation}
    \mathrm{Var}(\delta O_t)\propto \frac{1}{t}.
\end{equation}

Approximating the observable and score fluctuations as jointly Gaussian, the variance of their product satisfies
\begin{equation}
\mathrm{Var}(\delta O_tS_\theta)
=
\mathrm{Var}(\delta O_t)\mathrm{Var}(S_\theta)
+
\mathrm{Cov}(\delta O_t,S_\theta)^2
\leq
2\mathrm{Var}(\delta O_t)\mathrm{Var}(S_\theta).
\end{equation}

\begin{figure}[htbp]
\centering
\includegraphics[width=\textwidth]{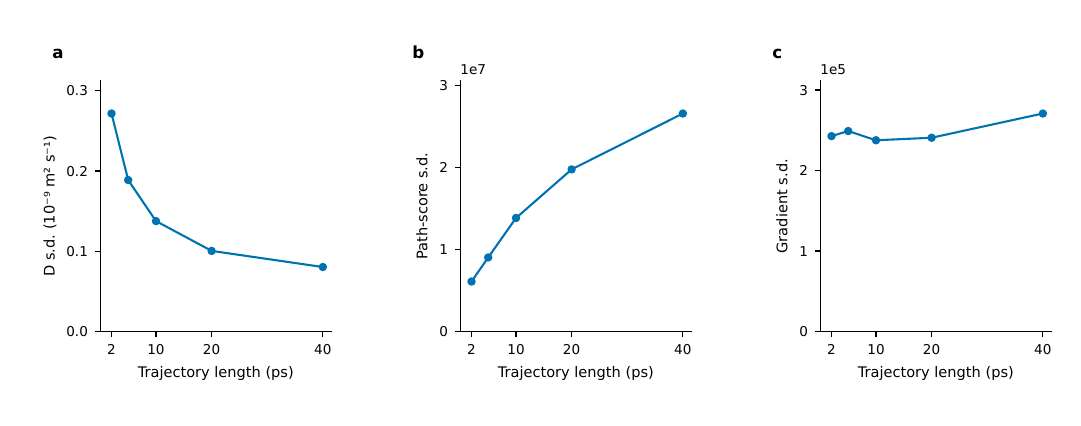}
\caption[Standard-deviation decomposition for the water gradient]{\textbf{Standard-deviation decomposition for the water gradient.}
(\textbf{a})~Sample standard deviation of diffusion coefficient across $128$ trajectories.
(\textbf{b})~Square root of the sample covariance trace of the accumulated dynamical path score.
(\textbf{c})~Standard deviation of the gradient estimate. 
}
\label{fig:water_variance_components}
\end{figure}

Therefore, the dependence on trajectory length is expected to cancel. Longer trajectories increase the variance of the accumulated path score but simultaneously improve the statistical estimate of the observable. 
We assessed this compensation for the initial NEP water model using $128$ separately equilibrated Langevin trajectories at $\gamma=1\,\mathrm{ps^{-1}}$ (Supplementary Fig.~\ref{fig:water_variance_components}).
The diffusion standard deviation decreases with trajectory length, while the cumulative path-score standard deviation increases. The resulting standard deviation of the gradient shows a modest variation in this range. 

It should be noted that the variance retains a dependence on the Langevin friction coefficient, which scales approximately as $O(1/\gamma)$. Because $\gamma$ is fixed throughout the simulations, this dependence does not affect the trajectory length scaling considered here. 

\clearpage
\section{Additional results of validating the water potential.}
\label{si:transport}

To validate the fine-tuned water potential, we selected four checkpoints along the fine-tuning trajectory at iterations $0$, $200$, $400$, and $680$. We further evaluated the initial (iteration $0$) and fine-tuned (iteration $680$) models at $288.15$ and $308.15\,\mathrm{K}$ to test temperature transferability.

For each model and temperature, five independent starting configurations were prepared by scaling the molecular positions to the cubic box length corresponding to the experimental density of liquid water at that temperature and atmospheric pressure~\citep{kell1975}, leaving intramolecular bond lengths and angles unchanged. All simulations use $512$ water molecules ($1536$ atoms), a time step of $\Delta t=1\,\mathrm{fs}$, and Langevin dynamics at $\gamma=1\,\mathrm{ps^{-1}}$ at fixed volume. At $288.15\,\mathrm{K}$, both the equilibration and production trajectories were lengthened to account for the slower decay of the stress autocorrelation in the initial model. 

\begin{figure}[t]
\centering
\includegraphics{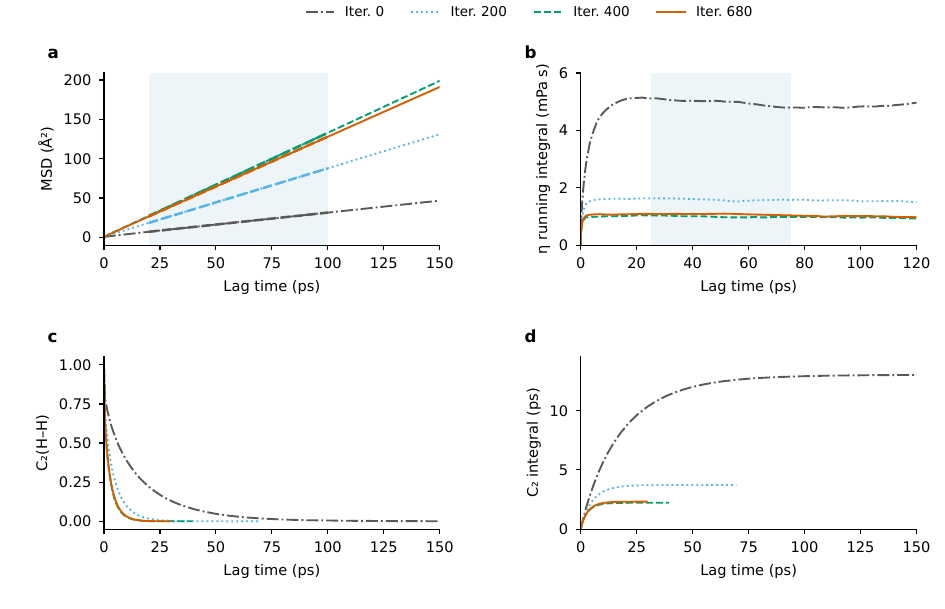}
\caption[Extraction of water transport coefficients]{\textbf{Extraction of water transport coefficients.}
(\textbf{a})~Mean MSD with the $20$--$100\,\mathrm{ps}$ fitting window
shaded and linear fits over that window.
(\textbf{b})~Mean viscosity running integral with its $25$--$75\,\mathrm{ps}$
plateau window shaded.
(\textbf{c})~Mean H--H second-rank rotational correlation and
(\textbf{d})~its running trapezoidal integral over each set's sampled window. Curves are
averages over five independent trajectories per parameter set.}
\label{fig:S2}
\end{figure}

Supplementary Fig.~\ref{fig:S2} displays the underlying correlation functions and extraction windows at $298.15\,\mathrm{K}$, averaged over the five independent production trajectories for each checkpoint. Supplementary Table~\ref{tab:S1} lists the three transport coefficients for the four checkpoints along the optimization iterations, and Supplementary Table~\ref{tab:S2} compares the initial and fine-tuned models across the three temperatures $288.15$, $298.15$ and $308.15\,\mathrm{K}$. Every entry is the mean and sample standard deviation over five independent production trajectories with the simulation lengths specified in Table~\ref{tab:water-validation}.

To quantify the temperature dependence of water dynamics, we analyzed the three dynamical observables using the Arrhenius relation
\begin{equation}
    Q(T) = Q_0 \exp\left(\pm\frac{E_a}{RT}\right),
\end{equation}
where $Q$ denotes the self-diffusion coefficient $D$, shear viscosity $\eta$, or second-rank rotational relaxation time $\tau_2$ of the intramolecular H--H vector, $R$ is the gas constant, and $E_a$ is the activation energy. The negative sign applies to $D$, while the positive sign applies to $\eta$ and $\tau_2$.

\begin{figure}[t]
\centering
\includegraphics{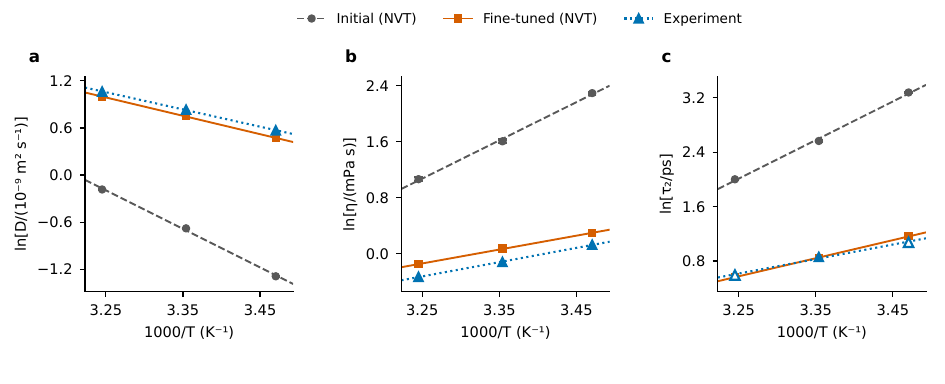}
\caption[Arrhenius analysis of water dynamics]{\textbf{Arrhenius analysis of water dynamics over 288.15--308.15 K.}
(\textbf{a})~Diffusion coefficient $D$, (\textbf{b})~shear viscosity $\eta$ and
(\textbf{c})~second-rank rotational relaxation time $\tau_2$ of the intramolecular H--H vector, for the initial and fine-tuned models and experimental measurements. Results are plotted as $\ln(Q/Q_{\mathrm{ref}})$ versus $1000/T$, where $Q$ represents $D$, $\eta$, or $\tau_2$. Simulation symbols represent means over five independent NVT trajectories. The corresponding standard errors are too small for the error bars to be visible in the figure. 
Lines represent linear Arrhenius fits, from which the apparent activation energies reported in Table~\ref{tab:Ea} are determined.}\label{fig:arrhenius}
\end{figure}

We performed ordinary least-squares fits of $\ln(Q/Q_{\mathrm{ref}})$ versus $1000/T$ at 288.15, 298.15 and 308.15 K, using the experimental density at each temperature. The resulting apparent activation energies, shown in Fig.~\ref{fig:arrhenius} and summarized in Table~\ref{tab:Ea}, characterize the temperature dependence
over this 20 K interval. Experimental activation energies were obtained using the same fitting procedure applied
to the reference data listed in Table~\ref{tab:S2}. Fine-tuning brings the apparent activation energies of all
three dynamical observables substantially closer to their experimental values, demonstrating improved temperature
dependence beyond the training condition.

\clearpage

\begin{table}[t]
\centering
\caption[Water validation protocol]{\textbf{Water validation protocol.} Production lengths are per
trajectory; five trajectories are used for each available parameter set.}
\label{tab:water-validation}
\begin{tabular}{lccc}
\toprule
$T$ (K) & Density (g\,cm$^{-3}$) & Equilibration & Production\\
\midrule
288.15 & 0.99910 & 400 ps & 10 ns\\
298.15 & 0.99704 & 200 ps & 5 ns\\
308.15 & 0.99403 & 200 ps & 5 ns\\
\bottomrule
\end{tabular}
\end{table}

\clearpage

\clearpage

\begin{table}[t]
\centering
\caption[Transport coefficients at $298.15\,\mathrm{K}$]{\textbf{Transport coefficients at $298.15\,\mathrm{K}$.}
$T=298.15\,\mathrm{K}$, $\rho=0.99704\,\mathrm{g\,cm^{-3}}$. Only $D$ and the
O--O RDF entered the loss. Values are means $\pm$ standard deviation over five independent
trajectories.}
\label{tab:S1}
\begin{tabular}{lccc}
\toprule
Iteration & $D$ ($10^{-9}$ m$^2$ s$^{-1}$) & $\eta$ (mPa\,s) & $\tau_2$ (ps) \\
\midrule
0   & $0.507\pm0.005$ & $4.98\pm0.28$ & $12.99\pm0.21$ \\
200 & $1.446\pm0.007$ & $1.59\pm0.08$ & $3.73\pm0.05$ \\
400 & $2.200\pm0.025$ & $0.99\pm0.12$ & $2.23\pm0.02$ \\
680 & $2.114\pm0.018$ & $1.08\pm0.07$ & $2.32\pm0.01$ \\
\midrule
Experiment & $2.299$ & $0.890$ & $2.36$ \\
\bottomrule
\end{tabular}
\end{table}

\clearpage

\begin{table}[t]
\centering
\caption[Transferability across temperature]{\textbf{Transferability across temperature.}
Initial and fine-tuned values at $288.15$, $298.15$ and $308.15\,\mathrm{K}$, each at the experimental density of that temperature. Experimental references: $D$~\citep{H2O_1}; $\eta$~\citep{korson1969}; $\tau_2$ at $298.15\,\mathrm{K}$~\citep{jonas1976,kumar2008}.}
\label{tab:S2}
\begin{tabular}{llccc}
\toprule
Quantity & Set & $288.15$ K & $298.15$ K & $308.15$ K \\
\midrule
\multirow{3}{*}{$D$ ($10^{-9}$ m$^2$ s$^{-1}$)}
  & Initial    & $0.275\pm0.005$ & $0.507\pm0.005$ & $0.832\pm0.011$ \\
  & Fine-tuned & $1.602\pm0.010$ & $2.114\pm0.018$ & $2.712\pm0.005$ \\
  & Experiment & $1.766$ & $2.299$ & $2.895$ \\
\midrule
\multirow{3}{*}{$\eta$ (mPa\,s)}
  & Initial    & $9.897\pm0.482$ & $4.977\pm0.283$ & $2.901\pm0.185$ \\
  & Fine-tuned & $1.346\pm0.079$ & $1.077\pm0.071$ & $0.861\pm0.077$ \\
  & Experiment & $1.138$ & $0.890$ & $0.720$ \\
\midrule
\multirow{3}{*}{$\tau_2$ (ps)}
  & Initial    & $26.51\pm0.75$ & $12.990\pm0.214$ & $7.396\pm0.057$ \\
  & Fine-tuned & $3.208\pm0.014$ & $2.318\pm0.011$ & $1.763\pm0.004$ \\
  & Experiment & $2.93^\ast$ & $2.36$ & $1.81^\ast$ \\
\bottomrule
\multicolumn{5}{p{15cm}}{\footnotesize 
$^\ast$The experimental $\tau_2$ values at $288.15$ and $308.15\,\mathrm{K}$ are digitized from the fitted temperature dependence of the intramolecular proton dipolar correlation time $\tau_d$ reported by Krynicki~\citep{krynicki1966}, and are therefore plotted as open symbols in Fig.~5c of the main text. Reading the same fitted curve at $298.15\,\mathrm{K}$ gives $2.28\,\mathrm{ps}$, consistent with the $2.36\,\mathrm{ps}$ value derived from the measurements of Jonas \emph{et al.}~\citep{jonas1976,kumar2008}. Separating the intermolecular contribution to the proton spin--lattice relaxation rate introduces an additional, model-dependent uncertainty in NMR estimates of $\tau_2$~\citep{lankhorst1982}. In the extreme-narrowing limit, the intramolecular correlation time inferred from proton relaxation corresponds to the time integral of $C_2(t)$, matching the definition of $\tau_2$ used here, whereas stretched-exponential (Kohlrausch--Williams--Watts) fit parameters and $1/e$ decay times characterize different relaxation measures.}
\end{tabular}
\end{table}

\begin{table}[t]
\centering
\caption[Apparent activation energies of water dynamics]{\textbf{Apparent activation energies of water dynamics over 288.15--308.15 K.}
Values for initial and fine-tuned models and experiments were obtained from three-point Arrhenius fits at $288.15$, $298.15$ and $308.15$~K (Fig.~\ref{fig:arrhenius}). Experimental values were fitted using the reference data listed in Table~\ref{tab:S2}. All activation energies are reported in $\mathrm{kJ\,mol^{-1}}$.}
\label{tab:Ea}
\begin{tabular}{lccc}
\toprule
Quantity & Initial & Fine-tuned & Experiment \\
\midrule
$D$ & $40.86$ & $19.44$ & $18.25$ \\
$\eta$ & $45.35$ & $16.47$ & $16.94$ \\
$\tau_2$ (HH) & $47.17$ & $22.11$ & $17.75$ \\
\bottomrule
\end{tabular}
\end{table}

\clearpage

\begin{table}[t]
\centering
\caption[Simulation and optimization parameters]{\textbf{Simulation and optimization parameters.}
$t$ is the trajectory length used to evaluate $D$, $\Delta t$ the time step,
$\gamma$ the Langevin friction, $T$ the thermostat temperature, $\rho$ the density,
$N_\mathrm{atoms}$ the number of atoms simulated, $K$ the
number of independent trajectories used to estimate the gradient, and $\alpha_D$, $\alpha_\mathrm{struct}$ the learning rates for the
dynamical and structural loss terms.}
\label{tab:params}
\begin{tabular}{lcc}
\toprule
Parameter & Molten Cu & water \\
\midrule
$t$ (ps)            & 1    & 2 \\
$\Delta t$ (ps)     & $10^{-3}$ & $10^{-3}$ \\
$\gamma$ (ps$^{-1}$) & 1   & 1 \\
$T$ (K)             & 1500 & 300 \\
$\rho$ (g\,cm$^{-3}$) & 7.94 & 0.994 \\
$D_\mathrm{target}$ ($10^{-9}$ m$^2$ s$^{-1}$) & $4.3^\dagger$ & $2.3^\ddagger$ \\
$N_\mathrm{atoms}$  & 256  & 1536 \\
$K$  & 64   & 64 \\
$\alpha_D$          & $10^{-4}$ & $10^{-3}$ \\
$\alpha_\mathrm{struct}$ & $10^{-6}$ & $10^{-2}$ \\
\bottomrule
\multicolumn{3}{l}{\footnotesize $^\dagger$Refs.~\citep{Cu0,Cu1,Cu2,Cu3};
$^\ddagger$Refs.~\citep{H2O_1,H2O_2,H2O_3}.}
\end{tabular}
\end{table}

\clearpage

\bibliography{references}